\documentclass[
superscriptaddress,
preprint,
 amsmath,amssymb,
 aps,
 prc
]{revtex4-1}

\usepackage{graphicx}
\usepackage{dcolumn}
\usepackage{bm}
\usepackage{soul,xcolor}
\usepackage{amssymb}
\usepackage{mathrsfs}
\usepackage[pdfencoding=auto,psdextra]{hyperref}
 \usepackage{CJK}

\setstcolor{blue}

\begin{document}
 \begin{CJK*}{UTF8}{}

\title{Thermodynamics of pairing transition for Odd-A nuclei}

\author{Tao Yan (\CJKfamily{gbsn}颜涛)}
\affiliation{School of Science, Jiangnan University, Wuxi 214122, China.}

 \author{Yanlong Lin (\CJKfamily{gbsn}林彦龙)}
\affiliation{School of Science, Jiangnan University, Wuxi 214122, China.}

 \author{Lang Liu (\CJKfamily{gbsn}刘朗)}
\email{liulang@jiangnan.edu.cn}
\affiliation{School of Science, Jiangnan University, Wuxi 214122, China.}

\begin{abstract}

The hot nucleus $^{171}$Yb is investigated by the covariant density functional theory with the PC-PK1 effective interaction.
The thermodynamic quantities are evaluated with the canonical ensemble theory.
The pairing correlations is treated by the shell-model-like approach, in which the particle numbers are conserved strictly and in which the blocking effect is handled exactly.
An S-shaped heat capacity versus temperature of $^{171}$Yb appears. It has been studied in terms of the blocking effect, the single-particle levels, the pairing gap, and defined seniority components, and compared to the heat capacity of $^{172}$Yb.
The pairing transition from the superfluid state to the normal state can result in the S-shaped heat capacity of $^{172}$Yb where the one-pair-broken and two-pair-broken states dominate, while the single-particle level structure near the Fermi surface is associated with the S-shaped heat capacity of $^{171}$Yb.
For odd-A nuclei, although the one-pair-broken and two-pair-broken states still contribute, the pairing gap and the pairing transition is relatively weak. 
The S-shaped heat capacity could be affected due to the blocking of the single-particle level near the Fermi surface.

\end{abstract}

\maketitle
\end{CJK*}

\section{Introduction}

The hot many-body systems in which the excited states are thermally distributed among particles have attracted widespread attention in nuclear physics.
In particular, the study of thermodynamics is essential for compound nuclei, heavy ion collisions and induced fusions~\cite{Bohr1998NuclearStructure:I,Weisskopf1937PR52:295--303,Bethe1937RMP9:69--244,Bertsch1988PR160:189--233,Hofmann1997PR284:137--380}.
Studies have shown that the pairing correlation plays a very important role in these phenomena and other nuclear thermodynamic properties~\cite{Dean2003RMP75_607--656}, such as the shape transition in hot nuclei~\cite{Zhang2017PRC96:054308,Zhang2018PRC97:054302}, the phase diagram structure of liquid-gas phase transition~\cite{Yang2019PRC100:054314,Yang2021PRC103:014304}, and the fragments produced in spallation reactions~\cite{Niu2018CPC42:034102}.

Thanks to the accurate measurement of energy level density~\cite{Melby1999PRL83:3150--3153,Guttormsen2003PRC68:034311}, S-shaped curves of heat capacity have been found in $^{161,162}$Dy, $^{171,172}$Yb~\cite{Schiller2001PRC63:021306R} and $^{166,167}$Er~\cite{Melby2001PRC63:044309}, and is explained as a superfluid-to-normal state transition.
The theoretical calculation and investigation for the nature of the S-shaped curve for heat capacity and other thermodynamic properties have also been performed in the nuclear shell model~\cite{Rombouts1998PRC58:3295--3304,Liu2001PRL87:022501,Langanke2005NPA757:360--372}, mean field model~\cite{Egido2000PRL85:26--29,Agrawal2000PRC62:044307,Sandulescu2000PRC61:044317,Niu2013PRC88:034308,Li2015PRC92:014302} and other models~\cite{Guttormsen2001PRC63:044301,Guttormsen2001PRC64:034319}.
In the mean field theory, one could define the superfluid and the normal-fluid phases for nuclei with BCS theory or the Bogoliubov transformation.
Clear pairing phase transitions has been obtained by the fact that the S shapes of heat capacity can be reproduced by most mean-field calculations including finite-temperature BCS~\cite{Sano1963PTP29:397--414,Goodman1981NPA352:30--44,Gambacurta2013PRC88:034324}, finite-temperature HFB with a pairing-plus-quadrupole Hamiltionian~\cite{Goodman1986PRC34:1942--1949}, as well as the self-consistent mean-field models in the non-relativistic~\cite{Reiss1999EPJA6:157--165} and relativistic framework~\cite{Niu2013PRC88:034308,Li2015PRC92:014302}.
The critical temperature for the pairing phase transition can be then derived in these models.

However, the particle number conservation is violated during the transition of pairing in the BCS theory and the Bogoliubov transformation.
The number conservation effects on the nuclear heat capacity have been investigated through the particle-number projection methods based on the finite-temperature BCS or HFB approaches~\cite{Esebbag1993NPA552:205--231,Esashika2005PRC72:044303,Gambacurta2013PRC88:034324}.
One of the consequences of this symmetry breaking is that there is a non-physical mutations in thermodynamic quantities, e.g., heat capacity and pairing gap, near the critical temperature.
These changes are unlikely to occur in a system with finite particle number and can be smoothed out after the particle number projection is introduced~\cite{Esashika2005PRC72:044303}.

Moreover, in the mean field theory with the BCS equation or the Bogoliubov transformation, the blocking effect is difficult to deal with. Therefore, researches on odd-A nuclei are quite rare. However, the shell-model-like approach (SLAP) can accurately treat the blocking effect, as well as keep the particle number conserved strictly~\cite{Zeng1983NPA405:1--28,Zeng1994PRC50:1388--1397,Meng2006FPC1:38--46}.
The SLAP has been used successfully for describing various phenomena concerning pairing correlations, e.g., the odd-even differences in moments of inertia~\cite{Zeng1994PRC50:746756}, the identical bands~\cite{Liu2002PRC66:024320,He2004EPJA23:217222}, the nuclear pairing phase transition~\cite{Wu2011PRC83:034323}, the antimagnetic rotation~\cite{Zhang2013PRC87:054314,Zhang2016PRC94:034305}, and the high-K rotational bands in the rare-earth~\cite{Liu2004NPA735:7785,Zhang2009NPA816:1932,Zhang2009PRC80:034313,Zhang2016SCPMA59:672012,Zhang2016NPA949:2234} and actinide nuclei~\cite{He2009NPA817:4560,Zhang2011PRC83:011304,Zhang2012PRC85:014324,Zhang2013PRC87:054308}, etc. 
At the same time, with the development of the covariant density functional theory (CDFT)~\cite{Ring1996PPNP37:193--263,Meng2006PPNP57:470--563,Vretenar2005PR409:101--259,Meng2021AB31:}, a successful point coupling interaction, PC-PK1, is proposed and proved to be very successful in the description of infinite nuclear matter and finite nuclei including the ground-state and low-lying excited states~\cite{Zhao2010PRC82_54319--54319}.
It has also been widely used in describing the spectroscopies of rod- and pear-shaped nuclei~\cite{Zhao2018IJMPE27:1830007}, the Coulomb displacement energies between mirror nuclei~\cite{Sun2011SCPMA54:210214}, quadrupole moments~\cite{Zhao2014PRC89:011301,Yordanov2016PRL116:032501}, magnetic and antimagnetic rotations~\cite{Zhao2011PLB699:181186,Zhao2011PRL107:122501,Meng2013FP8:5579,Peng2015PRC91:044329,Meng2016PS91:053008}, nuclear masses~\cite{Zhao2012PRC86:064324,Lu2015PRC91:027304,Xia2018DNDT121122:1215}, nuclear shape phase
transitions~\cite{Quan2018PRC97:031301}, chiral rotations~\cite{Zhao2017PLB773:1}, superheavy nuclei~\cite{Zhang2013PRC88:054324,Lu2014PRC89:014323,Agbemava2015PRC92:054310,Li2015FP10:102101}, etc.
The SLAP method should be combined with CDFT with PC-PK1 interaction, just as they have been successfully used to study the antimagnetic rotation of nuclei~\cite{Liu2019PRC99_024317},  the band crossing and shape evolution in $^{60}$Fe~\cite{Shi2018PRC97:034317}, the superdeformed rotational band in $^{40}$Ca~\cite{Wang2020PRC102:014321}, and to study the properties of hot nuclei.

In our previous work~\cite{Liu2015PRC92:044304}, thermodynamic properties of even-even nuclei have been studied within CDFT with the meson exchange. Heat capacity, entropy and level density are studied microscopically in terms of defined seniority components.
In this work, we took $^{171}$Yb as an example to study thermodynamics of pairing transition for odd-A nuclei within the CDFT+SLAP with PC-PK1 effective interaction. In addition, we also calculated $^{172}$Yb as a comparison.

This paper is organized as follows. A brief introduction to the cranking CDFT-SLAP theoretical framework is introduced in Sec.~\ref{sec:theory}. In Sec.~\ref{sec:nd}, the numerical details are presented. The results and discussion are given in Sec.~\ref{sec:res}. The last section is the summary.

\section{Theoretical framework}
\label{sec:theory}

The theoretical framework of point coupling CDFT has been introduced in details in Ref.~\cite{Zhao2010PRC82_54319--54319}.
Only a simple outline is given here.
A Lagrangian density is the starting point of CDFT, from which the Dirac equation for nucleons with local scalar $S(\bm{r})$ and vector $V^{\mu}(\bm{r})$ potentials can be derived as
\begin{eqnarray}\label{DiracE}
   \left[ \gamma_{\mu} (i \partial^{\mu} - V^{\mu}) - (m + S) \right] \psi_{\xi} = 0,
\end{eqnarray}
where
\begin{eqnarray}
   S(\bm{r}) = \Sigma_{S}, \ \  V(\bm{r}) = \Sigma^{\mu} + \vec{\tau} \cdot \vec{\Sigma}^{\mu}_{TV},
\end{eqnarray}
and $\psi_{\xi}$ is Dirac spinor.
In the above formula, the nucleon scalar-isoscalar $\Sigma_{S}$, vector-isoscalar $\Sigma^{\mu}$, and vector-isovector $\vec{\Sigma}^{\mu}_{TV}$ self-energies can be obtained by the various densities as follows
\begin{align}
    &\Sigma_{S}=\alpha_S\rho_S+\beta_S\rho^2_S+\gamma_S\rho^3_S+\delta_S\Delta\rho_S,\notag \\
    &\Sigma^{\mu}=\alpha_V j^{\mu}_V+\gamma_V ({j^{\mu}_V})^3+\delta_V\Delta j^{\mu}_V + e A^{\mu},\notag \\
    &\vec{\Sigma}^{\mu}_{TV}=\alpha_{TV} \vec{j}\,^{\mu}_{TV} + \delta_{TV} \Delta \vec{j}\,^{\mu}_{TV},
\end{align}
where the local densities and currents are defined as
\begin{align}\label{eq-density-curr}
    \rho_S(\bm r)&=\sum_{\xi}n_{\xi}\bar\psi_{\xi}(\bm r)\psi_{\xi}(\bm r), \notag \\
    {j}_V(\bm r)&=\sum_{\xi}n_{\xi}\bar\psi_{\xi}(\bm r)\gamma^{\mu}\psi_{\xi}(\bm r),\notag \\
    \vec{j}\,^{\mu}_{TV}(\bm r)&=\sum_{\xi}n_{\xi}\bar\psi_{\xi}(\bm r)\vec\tau\gamma^{\mu}\psi_{\xi}(\bm r), 
\end{align}
and $n_{\xi}$ is the occupation probability for single particle state $\xi$.
The iterative solution of these equations yields the total energy, quadrupole moments, single-particle energies, etc. 

In the SLAP method, the pairing correlation is then treated by diagonalizing the following Hamiltonian in the multi-particle configurations (MPCs) space, which are constructed by the single-particle levels obtained by the above CDFT solution. 
\begin{eqnarray}
  \label{eq:h}
  H &=& H_{\rm s.p.} + H_{\rm pair} \cr
    &=& \sum\limits_{i}\varepsilon_{i}a^{+}_{i}a_{i}
  -G\sum\limits^{i\neq j}_{i,j>0} a^{+}_{i}a^{+}_{\bar{i}}a_{\bar{j}}a_{j},
\end{eqnarray}
where $\varepsilon_{i}$ is the single-particle energy obtained from the Dirac equation (\ref{DiracE}), $\bar i$ is the time-reversal state of $i$, and $G$ represents constant pairing strength.
For a system with an even particle number $N = 2n$, the MPCs could be constructed as follows:
\begin{enumerate}
\item fully paired configurations (seniority $s = 0$):
\begin{eqnarray}
|c_1\bar{c}_1\cdots c_n\bar{c}_n\rangle=a^+_{c_1}a^+_{\bar{c}_1}\cdots a^+_{c_n}a^+_{\bar{c}_n}|0\rangle;
\end{eqnarray}
\item configurations with two unpaired particles (seniority $s = 2$)
\begin{eqnarray}
|i\bar{j}c_1\bar{c}_1\cdots c_{n-1}\bar{c}_{n-1}\rangle=a^+_{i}a^+_{\bar{j}}a^+_{c_1}a^+_{\bar{c}_1}\cdots a^+_{c_{n-1}}a^+_{\bar{c}_{n-1}}|0\rangle\quad\quad(i\ne j);
\end{eqnarray}
\item configurations with more unpaired particles (seniority $s=4, 6, \ldots$), see, e.g., Refs.~\cite{Zeng1983NPA405:1--28,Meng2006FPC1:38--46}. 
\end{enumerate}
In the case of odd nucleons system, one only need to block the state occupied by odd nucleons. For example, the MPCs for $s=0$ states can be expressed as, 
\begin{eqnarray}
|c_{b}c_1\bar{c}_1\cdots c_n\bar{c}_n\rangle=a^{+}_{b}(a^+_{c_1}a^+_{\bar{c}_1}\cdots a^+_{c_n}a^+_{\bar{c}_n})|0\rangle,
\end{eqnarray}
where $b$ denotes the single particle level blocked by the odd nucleon.

The Hamiltonian (\ref{eq:h}) have the good quantum numbers of the parity $\pi$ and the seniority $s$. As a result,  the MPC space can be written as:
\begin{eqnarray}
{\rm MPC~space} &=& (s=0,\pi=+) \oplus (s=0, \pi=-) \oplus \nonumber \\
                 && (s=2,\pi=+)  \oplus  (s=2, \pi=-) \oplus \nonumber \\ 
                 && \cdots
\end{eqnarray}
In practical calculations, the MPC space has to be truncated with an energy cutoff $E_c$, i.e., the configurations with energies $E_m-E_0 \leq E_c$ are used to diagonalize the Hamiltonian~(\ref{eq:h}), where $E_m$ and $E_0$ are the energies of the $m$th configuration and the ground-state configuration, respectively. 

After the diagonalization of the Hamiltonian (\ref{eq:h}), one could obtain the nuclear many-body wave function
\begin{eqnarray}
|\psi_\beta\rangle&=&\sum\limits_{c_{1}\cdots c_{n}}{v_{\beta,\,c_1\cdots c_n}}|c_1\bar{c}_1\cdots c_n\bar{c}_n\rangle  \cr
&& +\sum\limits_{i,j}{\sum\limits_{c_{1}\cdots c_{n-1}}{v_{\beta(ij),\,c_1\cdots c_{n-1}}}|i\bar{j}c_1\bar{c}_1\cdot\cdot\cdot c_{n-1}\bar{c}_{n-1}\rangle} \cr
&& + \cdots,
\end{eqnarray}
where $\beta = 0$ for the ground state, and $\beta = 1, 2, 3, \ldots$ for the excited states with the excitation energy $E_{\beta}$. $v_{\beta}$ means the coefficient after diagonalization.
The pairing energy and the pairing gap then can be calculated by~\cite{Meng2006FPC1:38--46,Canto1985PLB161:21--26,Egido1985PLB154:1--5,Shimizu1989RMP61:131--168}
\begin{align}
E_{\rm pair} &= \langle\psi_{\beta}\mid{H_{\rm pair}}\mid\psi_{\beta}\rangle, \\
\Delta_{\beta}&=G\left[ -\frac{1}{G} \langle \Psi_{\beta} | H_{\rm p} | \Psi_{\beta} \rangle \right]^{1/2}.
\end{align}

By assuming the hot many-body system is a canonical ensemble~\cite{Sumaryada2007PRC76:024319}, the nuclear thermodynamic properties can be calculated with the excited energies $E_{\beta}$ and the energy density $\eta(E_{\beta})$ in the SLAP.
The canonical partition function $Z$, average energy $\langle E \rangle$, heat capacity $C_V$, and entropy $S$ are defined with the following equations,
\begin{eqnarray}
\label{eq:z}
Z &=& \sum\limits^{\infty}_{\beta=0}\eta(E_{\beta})\,e^{-E_{\beta}/T}, \\
\label{eq:e}
\langle E \rangle &=& Z^{-1}\sum\limits^{\infty}_{\beta=0}E_{\beta} \,\eta(E_{\beta})\,e^{-E_{\beta}/T}, \\
\label{eq:cv}
C_V &=& \frac{\partial \langle E \rangle}{\partial T},  \\
\label{eq:s}
S &=& \frac{\partial\,(T\ln Z)}{\partial T} = \frac{\langle E(T) \rangle}{T} + \ln Z,
\end{eqnarray}
where $E_{\beta}$ is the excitation energy which can be obtained from the SLAP method with CDFT, and the corresponding level density $\eta (E_{\beta})$ is taken as $2^s$, i.e., the degeneracy of each state. By means of the partition function, one can also evaluate the ensemble average pairing gap energy as
\begin{eqnarray}\label{eq:gap}
\tilde{\Delta} = Z^{-1}\sum\limits^{\infty}_{\beta=0} \Delta_{\beta} \,\eta(E_{\beta})\,e^{-E_{\beta}/T}.
\end{eqnarray}

\section{Numerical details}
\label{sec:nd}

In this work, CDFT is used to calculate the single-particle levels of $^{171}$Yb and $^{172}$Yb self-consistently. The diagonalization of the Hamiltonian including the pairing force in Eq.~(\ref{eq:h}) can be performed in the SLAP MPCs constructed by single-particle levels. The thermodynamic quantities are evaluated by using Eqs.~(\ref{eq:z}-\ref{eq:gap}) with the excitation energy and the level density after diagonalization. Besides, in order to study blocking effects and detailed single-particle features, we create $^{171}$Yb* artificially as a bridge between $^{171}$Yb and $^{172}$Yb. The neutron excitation spectrum of  $^{171}$Yb* is obtained by diagonalization in the SLAP MPC space which is constructed by the CDFT single-particle levels of $^{172}$Yb with one artificially removed neutron. The proton excitation spectrum of $^{171}$Yb* is the same as the proton excitation spectrum of $^{172}$Yb. The axially deformed harmonic oscillator basis with $20$ major shells is adopted to solve the Dirac equation~(\ref{DiracE})~\cite{Ring1997CPC105:77--97}.
The effective interaction is chosen as PC-PK1~\cite{Zhao2010PRC82_54319--54319}.
In the construction of the multi-particle configurations, 15 single-particle levels around Fermi surfaces and $8$ pairs of valence particles are included for neutrons and protons, respectively.
This also indicates that the highest seniority number is 16.
The energy level span for neutrons above and below the Fermi surface is chosen as 3.5 MeV, while that for protons is chosen as 5.2 MeV.  
The effective pairing strengths can, in principle, be determined by the experimental odd-even differences in the nuclear binding energies, and are connected with the dimension of the truncated MPC space. The odd-even mass difference is defined, e.g., for neutron, as:
\begin{eqnarray}
\Delta_n = B(N,Z) - \dfrac{1}{2}[ B(N-1,Z)+B(N+1,Z) ],
\end{eqnarray}
where the $B(N,Z)$ is the binding energy of the nucleus with neutron number $N$ and proton number $Z$.
The pairing strength in our calculations for neutron $G_{\rm n}$ is fixed to 0.19 MeV, and $G_{\rm p}$ for proton is 0.22 MeV with $E_c$=30 MeV.
The convergency of calculated results with energy cut off $E_c$ is checked.  
We find that the calculated heat capacities with $E_c = 20, 25$, and 30 MeV obtain consistent values from 0 to 1 MeV. In the following calculations, the energy cutoff $E_c$ is fixed as 30 MeV. The dimensions of the neutron and proton MPC spaces are about $2\times10^6$ and $6\times10^5$ for $^{171}$Yb, $2\times10^6$ and $6\times10^5$ for $^{171}$Yb*, and $6\times10^5$ and $6\times10^5$ for $^{172}$Yb, respectively.

\section{Results and discussion}
\label{sec:res}
\subsection{Thermodynamics}
\begin{figure}[ht!]
\centering
 \includegraphics[scale=.6]{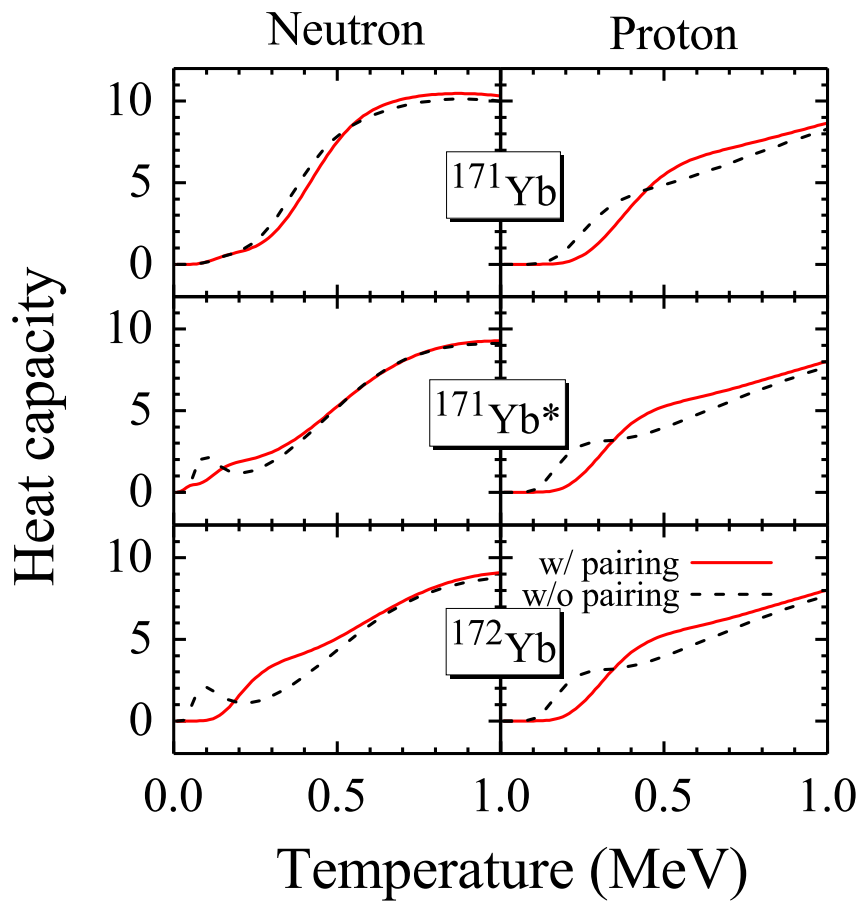}
 \caption{(Color online). Neutron (the left column) and proton (the right column) heat capacities for $^{171}$Yb (the top row), $^{171}$Yb* (the middle row), and $^{172}$Yb (the bottom row) as functions of temperature with (red solid lines) and without pairing (black dashed lines). }
\label{fig:cv}
\end{figure}

The heat capacities of neutron (the left column) and proton (the right column) for $^{171}$Yb (the top row), $^{171}$Yb* (the middle row), and $^{172}$Yb (the bottom row) as functions of temperature with (red solid line) and without (black dashed line) pairing are shown in Fig.~\ref{fig:cv}.
$^{171}$Yb* is explained in Sec.~\ref{sec:nd}.
The first column and the second column show the neutron and proton heat capacity, respectively.
It can be found that $^{172}$Yb presents the S-shaped heat capacity curve with the pairing force.
The S-shaped curve of $^{172}$Yb starts to occur at low temperatures ($T\sim 0.17$ MeV), and exhibits the second turning point around $T\sim 0.3$ MeV.
After that, it increases almost linearly at high temperatures. 
The microscopic mechanism of the S-shaped heat capacity of $^{172}$Yb as an even-even nucleus is similar to that of $^{162}$Dy, which has been discussed in details in previous work~\cite{Liu2015PRC92:044304}.
Moreover, the behavior of heat capacity curve for $^{172}$Yb given by our calculation is very consistent with the experimental extraction~\cite{Schiller2001PRC63:021306R}.
However, it is interesting to find that the S shape even appears in the heat capacity of $^{171}$Yb neutrons.
One can find that the first inflection point of the S-shaped curve of $^{171}$Yb neutrons appears at $T\sim 0.3$ MeV, and the second inflection point is presented around $T\sim 0.6$ MeV.
Also, unlike $^{172}$Yb, there is no linear increase at high temperatures in $^{171}$Yb neutrons.
Meanwhile, we carried out the calculation of $^{171}$Yb* which is obtained by the single-particle levels of $^{172}$Yb with one artificially removed neutron. It is found that the heat capacity of $^{171}$Yb* neutrons has no obvious S shape compared with $^{171}$Yb. The only difference between $^{171}$Yb and $^{171}$Yb* is the single-particle level structure, and it probably leads to the different behaviors of the heat capacity.

The heat capacity curve of $^{171}$Yb neutrons without pairing tends to coincide with the case with pairing, and $^{171}$Yb* shows the same behavior at $T\geq 0.4$ MeV.
They are both odd-neutron systems, so the effect of pairing correlations is relatively weak, and our calculation results have also confirmed this. Same as $^{171}$Yb*, the heat capacity curve of $^{172}$Yb has a small fluctuation at $T\sim 0.1$ MeV when there is no pairing force, and then it rises almost linearly, which is very analogous to the results of the pure Fermi gas model~\cite{Schiller2001PRC63:021306R}. However, the heat capacity of $^{172}$Yb at $T\le 0.6$ MeV has a significant difference between the case with pairing and the case without pairing.
The proton curves of $^{171}$Yb* are identical to that of $^{172}$Yb because the same proton single-particle level is used. It can be found that, in the case with pairing, all three results show S-shaped heat capacity curves.
These are also in line with the results of our discussion for even-even nucleons system~\cite{Liu2015PRC92:044304}.
The heat capacities of $^{171}$Yb and $^{171}$Yb* without pairing grow almost linearly at high temperatures and have slight fluctuations at low temperatures.

\begin{figure}[ht!]
\centering
 \includegraphics[scale=.6]{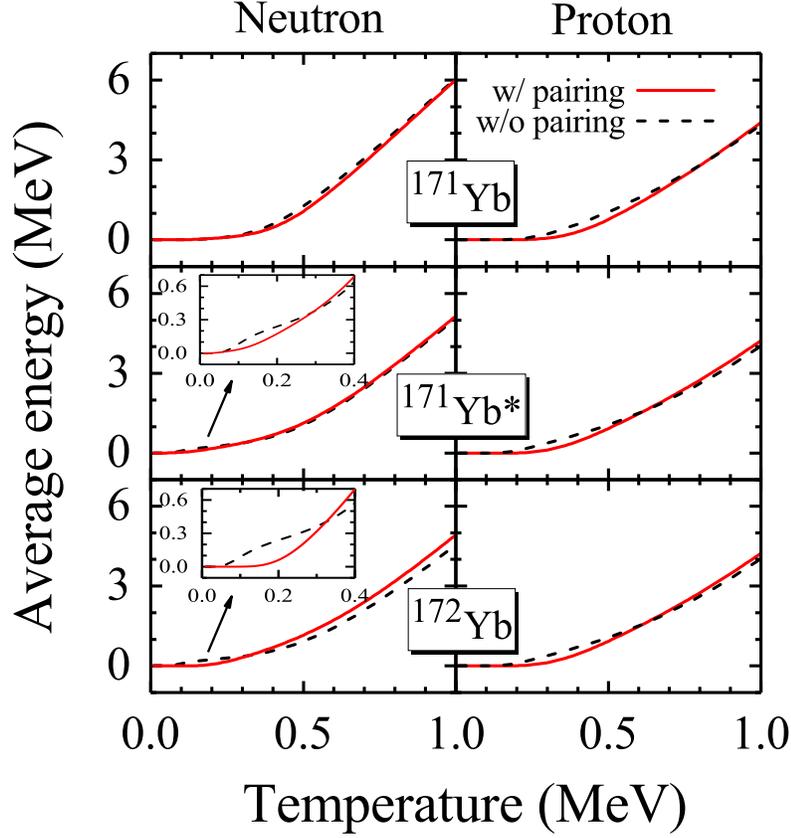}
 \caption{(Color online) Neutron (the left column) and proton (the right column) average excitation energies for $^{171}$Yb (the top row), $^{171}$Yb* (the middle row), and $^{172}$Yb (the bottom row) calculated with (red solid line) and without (black dashed line) pairing correlations as functions of temperature. The two inserting subfigures are the average excitation energies of $^{171}$Yb* and $^{172}$Yb at low temperature.}
\label{fig:e}
\end{figure}

The average excitation energies of neutrons and protons defined in Eq.~(\ref{eq:e}) for $^{171}$Yb (the top row), $^{171}$Yb* (the middle row), and $^{172}$Yb (the bottom row) calculated with (red solid line) and without (black dashed line) pairing correlations as functions of temperature are shown in Fig.~\ref{fig:e}. The first column and the second column are the average excitation energies of neutrons and protons, respectively. It can be seen that the average excitation energies of neutrons for $^{171}$Yb and $^{172}$Yb are approximately zero at low temperatures when the pairing correlation is considered. However, there are obvious changes around $T\sim 0.3$ MeV and $T\sim 0.17$ MeV for $^{171}$Yb and $^{172}$Yb respectively. Since the heat capacity is calculated as the partial derivative of the average energy with respect to temperature with Eq.~(\ref{eq:cv}), these also correspond to the first turning point of S-shaped heat capacity curve of neutrons for $^{171}$Yb and $^{172}$Yb in Fig.~\ref{fig:cv}.
Nevertheless, the average excitation energy of neutrons for $^{171}$Yb* starts to change when the temperature is very low, so its neutron heat capacity curve does not have an obvious suppression at low temperatures.

It can be seen from the second column of Fig.~\ref{fig:e} that the average excitation energies are about zero at $T\sim 0.3$ MeV, and then show a linear upward trend when the pairing correlations are taken into account among these three curves. In the absence of pairing correlations, all the average excitation energies increase linearly beyond $T\sim 0.2$ MeV. The linear increase of the average energy indicates that the number of the excited states increases uniformly with temperature.

\begin{figure}[ht!]
  \centering
    \includegraphics[scale=1.25]{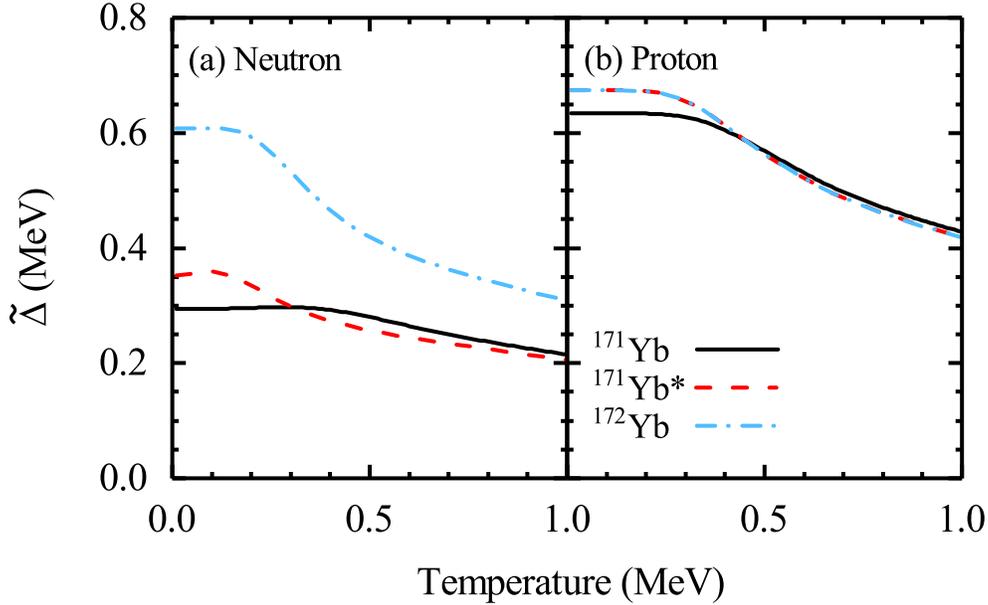}
  \caption{(Color online) The pairing gap of neutrons (a) and protons (b) for $^{171}$Yb (black solid lines), $^{171}$Yb* (red dashed line), and $^{172}$Yb (blue dash-dotted line) as functions of temperature calculated by SLAP.}
\label{fig:gap}
\end{figure}

\subsection{Pairing gap}
In order to further study the influence of pairing correlations on thermodynamic quantities, the pairing gaps of neutrons (a) and protons (b) are calculated with Eq.~(\ref{eq:gap}) and showed in Fig.~\ref{fig:gap}.
It can be found that the pairing gap of $^{172}$Yb changes very little at low temperatures, and begins to drop significantly after $T\sim 0.17$ MeV.
This temperature, where the pairing gap starts to change, exactly corresponds to the temperature of the first inflection point of the S-shaped heat capacity curve for $^{172}$Yb as shown in Fig.~\ref{fig:cv}.
The dropping of the pairing gap results in the increasing number of the Cooper-pair-broken excited states, and thus a rapid increase of the heat capacity.
The changes of the curve indicate the gradual transition of pairing correlations from a superfluid state to a normal state in the hot nucleus.
The second inflection point of heat capacity occurs at temperature about $0.3$~MeV, which is usually considered as the critical temperature for the phase transition of pairing correlations.
It locates at about 0.5$\tilde{\Delta}$(0)~\cite{Schiller2001PRC63:021306R}, where $\tilde{\Delta}$(0) is the pairing gap at zero temperature.
In our calculations, the temperature of $0.3$~MeV at which the second inflection point occurs is half of the pairing gap at zero temperature ($\sim 0.6$ MeV).
These properties are consistent with the even-even nuclei $^{162}$Dy discussed in Ref.~\cite{Liu2015PRC92:044304}.

However, in the case of odd-A nuclei, we can find that the magnitude of the pairing gaps for $^{171}$Yb and $^{171}$Yb* are about only half of $^{172}$Yb.
The pairing gap of $^{171}$Yb neutrons changes little with temperature before $T\sim 0.3$ MeV, and decreases almost linearly after that.
Although this temperature can correspond to the first inflection point of the heat capacity of $^{171}$Yb, it is difficult to find the temperature corresponding to the second inflection point of the heat capacity.
Moreover, the temperature ($T\sim 0.6$ MeV) of the second inflection point for the heat capacity does not correspond to the half of the pairing gap at zero temperature ($\tilde{\Delta}(0)\sim 0.3$ MeV).
For $^{171}$Yb*, the pairing gap of neutrons begins to drop significantly at $T\sim 0.1$ MeV, and has a similar trend with that of $^{171}$Yb at high temperature.
Since $^{171}$Yb* does not show an obvious S-shaped heat capacity curve, it is difficult to find the temperature corresponding to the pairing gap change.
In addition, it is found that their pairing gaps of $^{171}$Yb, $^{171}$Yb* and $^{172}$Yb protons, as even-particle systems, change with temperature can explain the behavior of their heat capacity curves of protons.
Hence, compared with even-even nuclei $^{172}$Yb, the CDFT+SLAP calculation seems to indicate that the heat capacity of $^{171}$Yb and $^{171}$Yb* heat capacity are not directly related to the phase transition of the pairing correlations, not to mention that the effect of pairing correlations is so weak.
Although the pairing gaps of $^{171}$Yb and $^{171}$Yb* behave similarly, the heat capacity curves are very different.
The only difference between them is the single-particle level.
Therefore, it is necessary for us to study their single-particle structure.

\begin{figure}[ht!]
  \centering
    \includegraphics[scale=.8]{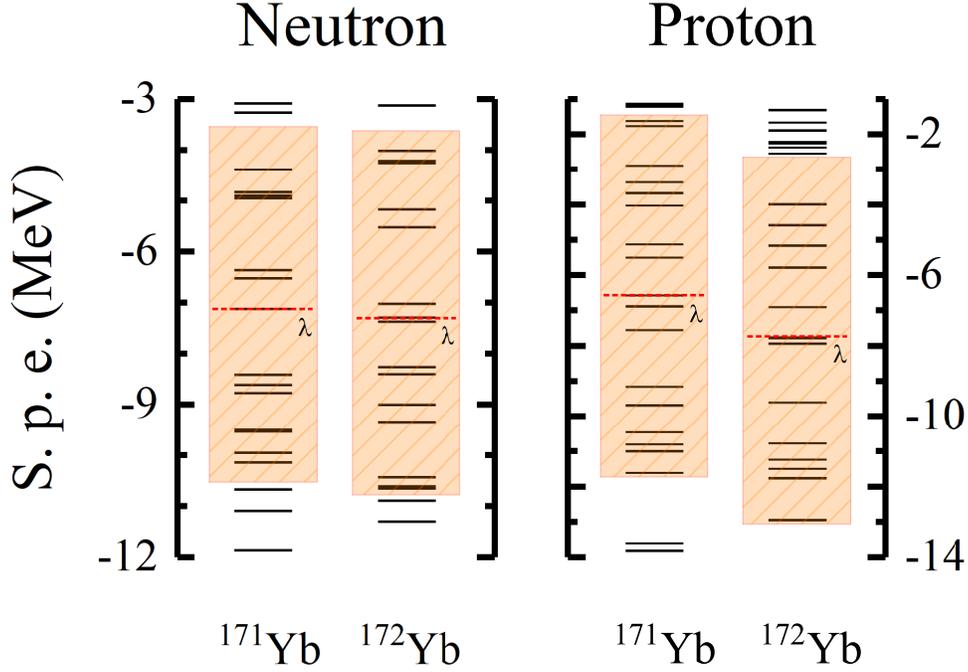}
  \caption{(Color online) The single-particle levels of neutrons and protons for $^{171}$Yb and $^{172}$Yb. The single-particle levels in shadows (orange area) are used to construct the SLAP MPCs. The red dashed lines labeled with $\lambda$ represent the Fermi surface.}
\label{fig:level}
\end{figure}

\subsection{Single-particle levels and blocking effects}

The single-particle levels of neutrons and protons for $^{171}$Yb and $^{172}$Yb are shown in Fig.~\ref{fig:level}. The left side is the single-particle levels of the neutrons for $^{171}$Yb and $^{172}$Yb. The red dashed lines labeled with $\lambda$ represent the Fermi surface. 
It can be found that, above and below the Fermi surface, especially below the Fermi surface, there is a relatively large energy gap.
That means that the single-particle level density for neutrons near the Fermi surface of $^{171}$Yb is smaller than that of $^{172}$Yb. Therefore, when the same temperature is raised, $^{171}$Yb is more difficult to generate pairing excitation than $^{172}$Yb.
It explains that $^{171}$Yb neutrons has a larger temperature scale where the curve is suppressed than other two curves at low temperatures in Fig.~\ref{fig:cv}.
The right side of Fig.~\ref{fig:level} is the single-particle levels of the protons for $^{171}$Yb and $^{172}$Yb. It can be seen that there are no obvious differences between the single-particle level structure for protons of $^{171}$Yb and $^{172}$Yb near the Fermi surface.
Therefore, it does not have a big impact on the heat capacity curve for protons.

\begin{figure}[ht!]
\centering
 \includegraphics[scale=.7]{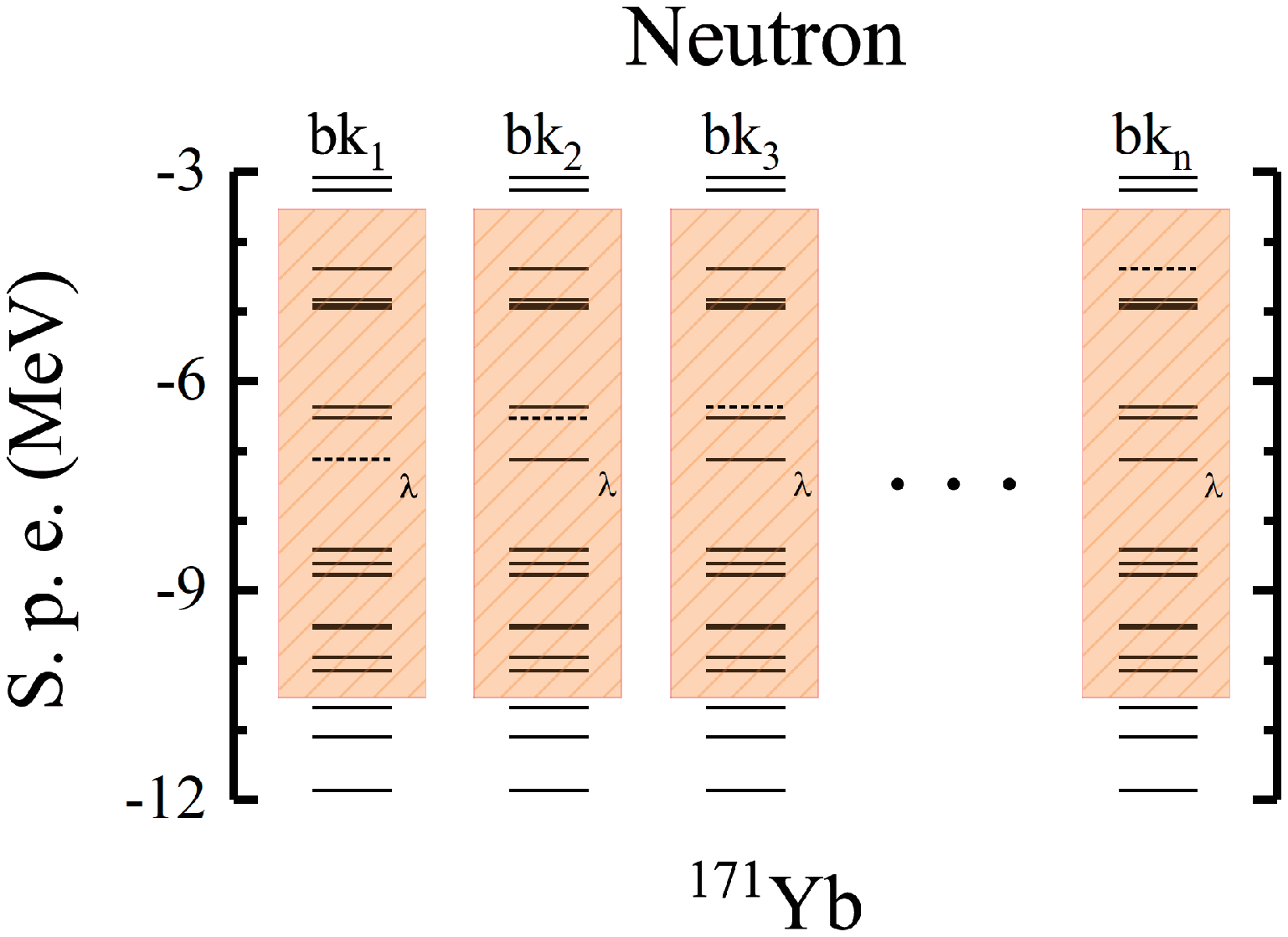}
 \caption{(Color online) The schematic diagram of blocking levels for neutrons of $^{171}$Yb. The single-particle levels in shadows (orange area) are used to construct the SLAP MPCs. $\lambda$ represents the Fermi surface. The dashed lines illustrate the blocking levels. ``bk$_{1}$" represents the case of blocking the Fermi surface; ``bk$_{2}$" represents the case of blocking the first level above the Fermi surface; ``bk$_{n}$" represents the case of blocking the $(n-1)$th level above the Fermi surface.}
 \label{fig:leblk}
\end{figure}

\begin{figure}[ht!]
\centering
 \includegraphics[scale=.6]{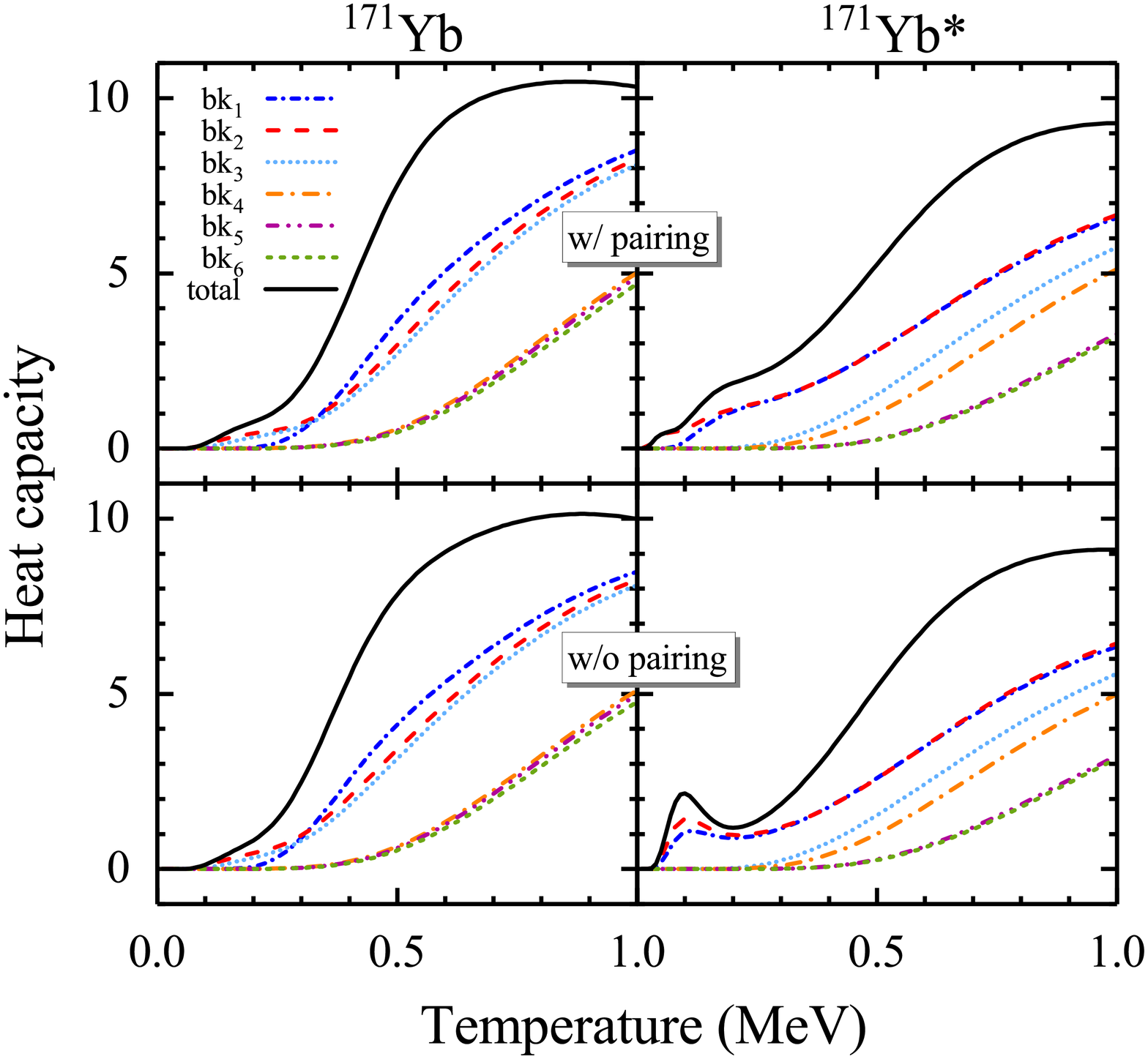}
 \caption{(Color online) The heat capacities of neutrons for $^{171}$Yb (the left column) and $^{171}$Yb* (the right column) with different levels blocked and total as functions of temperature with (the first row) and without (the second row) pairing. ``bk$_{1}$" represents the case of blocking the Fermi surface; ``bk$_{2}$" represents the case of blocking the first level above the Fermi surface; ``bk$_{6}$" represents the case of blocking the fifth level above the Fermi surface as illustrated in Fig.~\ref{fig:leblk}.}
 \label{fig:hcbk}
\end{figure}

In the odd nucleons system, it is very important to accurately deal with the blocking effects.
Figure~\ref{fig:hcbk} shows the heat capacities of neutrons for $^{171}$Yb (the left column) and $^{171}$Yb* (the right column) with different levels blocked and total as functions of temperature with (the first row) and without (the second row) pairing.
Among them, ``bk$_{1}$" represents the case of blocking the Fermi surface; ``bk$_{2}$" represents the case of blocking the first level above the Fermi surface; ``bk$_{n}$" represents the case of blocking the $(n-1)$th level above the Fermi surface. In our calculation, the maximum of $n$ is 6.
Figure~\ref{fig:leblk} schematically illustrates how blocking level is handled in SLAP.
It can be found that the total heat capacity curve with pairing of the neutrons for $^{171}$Yb mainly comes from the contribution of ``bk$_{1}$", ``bk$_{2}$", and ``bk$_{3}$", while ``bk$_{4}$" - ``bk$_{6}$" only make a numerical contribution to the total heat capacity and have little effects on the shape of the curve. The total heat capacity of the neutrons for $^{171}$Yb* mainly comes from the contribution of ``bk$_{1}$" and ``bk$_{2}$". ``bk$_{3}$" - ``bk$_{6}$" also only make a numerical contribution to the total heat capacity curve.
Similarly, for $^{171}$Yb, it can be seen that the total heat capacity curve without pairing of the neutrons also mainly comes from the contribution of ``bk$_{1}$", ``bk$_{2}$", and ``bk$_{3}$". In the case of $^{171}$Yb*, the total heat capacity of its neutrons is mainly derived from the contribution of ``bk$_{1}$" and ``bk$_{2}$".
Therefore, our calculations show that it is very important to correctly handle the blocking of the single-particle level near the Fermi surface when studying the thermodynamic properties of odd-A nuclei. 

\begin{figure}[ht!]
\centering
 \includegraphics[scale=.6]{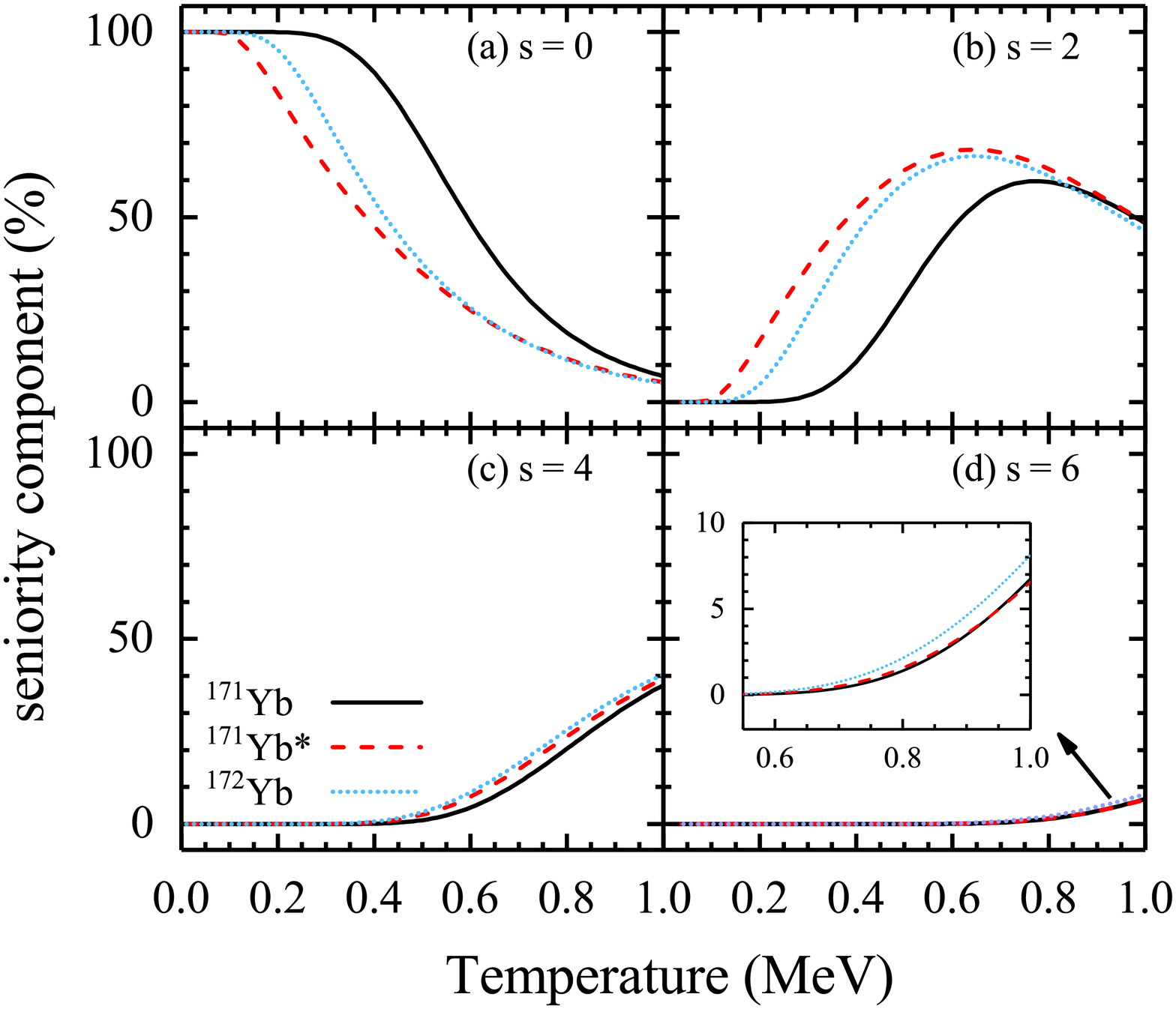}
 \caption{(Color online) The seniority components $\chi_{s}$ of neutron for $^{171}$Yb (black solid line), $^{171}$Yb* (red dashed line), and $^{172}$Yb (blue dotted line) with different seniority numbers $s = 0$ (a), $s = 2$ (b), $s = 4$ (c), and $s = 6$ (d) as functions of temperature. Insert: the locally enlarged subfigure with seniority numbers $s = 6$ at high temperatures.}
 \label{fig:sc}
\end{figure}

\subsection{Seniority component}
In order to provide a microscopic picture of the nuclear pairing transition, it is interesting to explore how many Cooper pairs would be broken with the increasing temperature in nuclei $^{171}$Yb, $^{171}$Yb*, and $^{172}$Yb. Here we follow the definition of the seniority component in Ref.~\cite{Liu2015PRC92:044304} 
\begin{equation}
  \chi_s = Z^{-1} \sum\limits_{\beta\in \{s\}} \eta(E_{\beta}) e^{-E_{\beta}/T}.
\end{equation}\label{s_com}
The seniority component just represents the contribution with each seniority number of the excited states.
Since SLAP method can accurately deal with blocking by removing the levels occupied by the odd nucleon from MPCs, we can study the seniority components of odd-A nuclei and even-even nuclei under the same framework.
In Fig.~\ref{fig:sc}, the seniority components $s$ of neutrons for $^{171}$Yb, $^{171}$Yb*, and $^{172}$Yb with different seniority numbers $s = 0$ (a), $s = 2$ (b), $s = 4$ (c), and $s = 6$ (d) are shown as functions of temperature. We could see that the $s = 0$ states contribute nearly 100\% below  $T\sim 0.3$ MeV, $0.1$ MeV, and $0.17$ MeV for $^{171}$Yb, $^{171}$Yb*, and $^{172}$Yb, respectively. This is again consistent with the vanishing heat capacity at low temperatures as shown in Fig.~\ref{fig:cv}.
With the  temperature $T\ge 0.3$ MeV,  $0.1$ MeV, and $0.17$ MeV respectively, the contribution of the $s = 0$ states falls down, while the contribution of the $s = 2$ states goes up. This corresponds to the first inflection point in the heat capacity curve of $^{171}$Yb and $^{172}$Yb.
However, the inflection point of the $s = 2$ states of $^{171}$Yb* does not have an accurate correspondence in the heat capacity curve, indicating once again that the pairing transition from the superfluid state to the normal state was suppressed by the blocking effect in $^{171}$Yb*.
Due to the influence of the single-particle level structure, the S-shaped heat capacity curve of $^{171}$Yb, which should have been smoothed out by the blocking effect, appears.
It is worth noting that the increasing rate of the contribution from the high excitation states with $s = 2$ of $^{171}$Yb starts to decrease at $T\sim0.6$ MeV. Thus, the increasing of corresponding seniority component becomes slower. This also corresponds to the second inflection point of the neutron heat capacity curve for $^{171}$Yb at $T\sim0.6$ MeV in Fig.~\ref{fig:cv}.
Similar to $^{171}$Yb, the increasing rate of the contribution from the high excited states with $s = 2$ of $^{172}$Yb starts to decrease at $T\sim0.3$ MeV. So the heat capacity curve of $^{172}$Yb has a second inflection point at $T\sim0.3$ MeV. Above $T\ge0.4$ MeV and $0.6$ MeV respectively, the contributions of $s = 4$ states and $s = 6$ states of $^{171}$Yb, $^{171}$Yb*, and $^{172}$Yb start to increase. The states with $s = 6$ have almost no contributions between $T\sim0$ MeV and $T\sim0.8$ MeV, and contribute less than 8\% at $T\sim1$ MeV. 

\begin{figure}[ht!]
\centering
 \includegraphics[scale=.5]{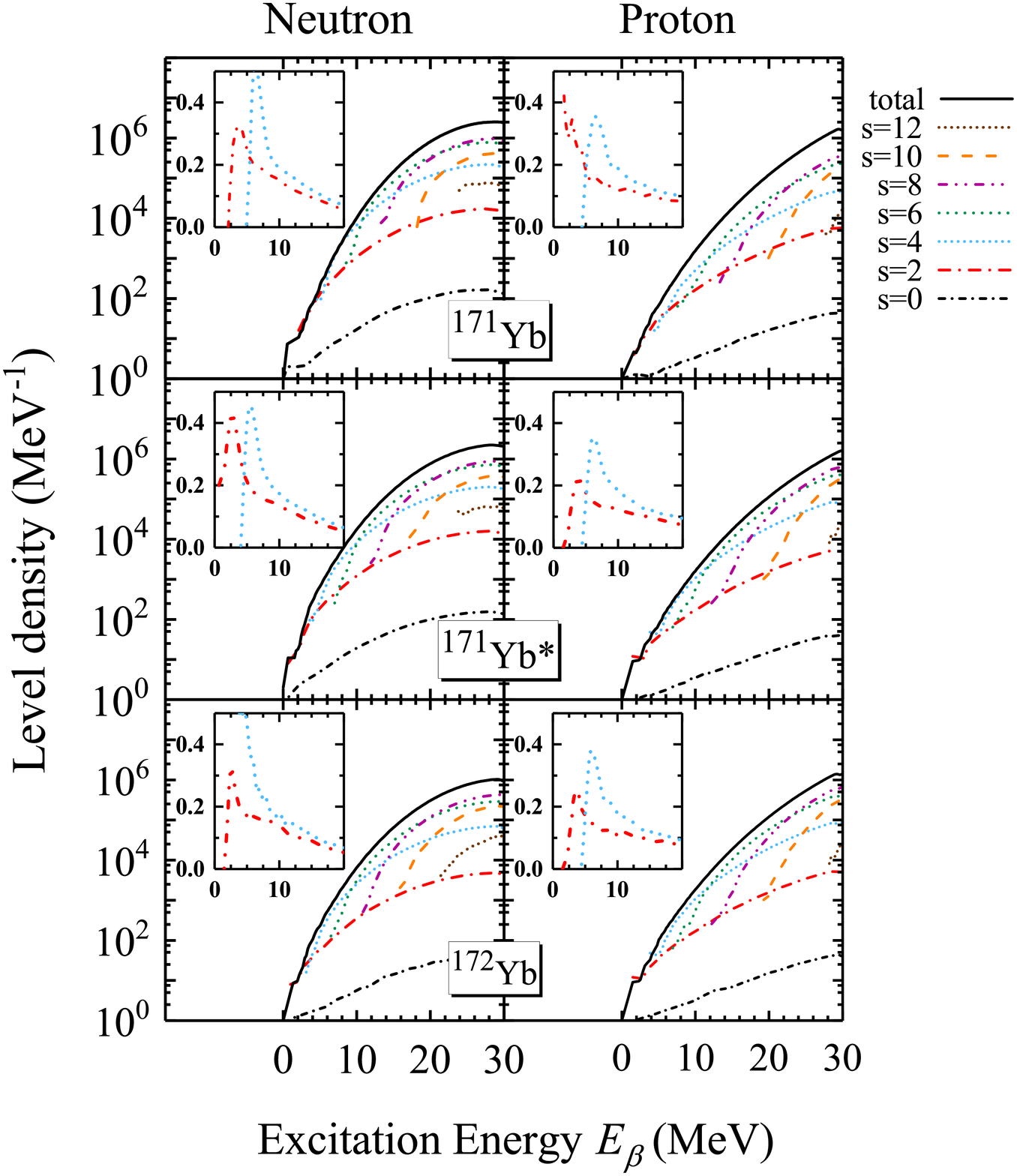}
 \caption{(Color online) Neutrons (the left column) and protons (the right column) level densities for $^{171}$Yb (the top row), $^{171}$Yb* (the middle row), and  $^{172}$Yb (the bottom row) with different seniority numbers $s = 0,2,4,6,8,10,12$ and the total contribution as functions of excitation energy. Insert: the slope of level density curves with different seniority numbers $s = 2,4$ as functions of excitation energy.}
 \label{fig:level_density}
\end{figure}

The level density has been introduced to characterize pairing transition in hot nuclei. In our calculations, the number of the excited states of nuclei can be obtained after the diagonalization of Hamiltonian [Eq.~(\ref{eq:h})].
It is nothing but the degeneracy of the excited states, and can be studied in terms of different seniority contributions. In Fig.~\ref{fig:level_density}, the neutron and proton level densities for $^{171}$Yb (the top row), $^{171}$Yb* (the middle row), and $^{172}$Yb (the bottom row) with $s = 0,2,4,6,8,10,12$ and the total contributions as functions of excitation energy are shown. Additionally, in order to reflect the fluctuations of energy level density more intuitively, the slope of level density curves with different seniority numbers $s = 2,4$ are displayed as representatives with insert figures.
It is worth noting that the level densities of $^{171}$Yb and $^{171}$Yb* neutrons are greater than that of $^{172}$Yb.
For odd-A nuclei, we are able to consider the blocking effect accurately. In a naive picture, different excited states can be configured by the same MPC and different blocking levels. Therefore, the number of excited states of odd-A nuclei is more than that of even-even nuclei.
In the previous work \cite{Liu2015PRC92:044304}, obvious protrusions can be seen in the level density curves with $s\neq 0$ states.
In our work, although no obvious protrusions can be seen on the plots of energy level density, the changing rate of level density with excitation energy ($E_{\beta}$) also fluctuates to a certain extent.
From the insert figures,
 we can see that the derivatives of the $s = 2$ and 4 level density curves for neutrons have peaks near $E_{\beta}\sim 4$ MeV and $E_{\beta}\sim 6$ MeV, respectively. It shows that the level density curves of $s = 2$ and 4 states change significantly with $E_{\beta}$ in the vicinity of $E_{\beta}\sim 4$ MeV and $E_{\beta}\sim 6$ MeV. Similarly, in the case of protons, the level density curves corresponding to the $s = 2$ and 4 states for protons vary considerably around $E_{\beta}\sim 3$ MeV and $E_{\beta}\sim 7$ MeV. The peaks at 4 MeV ($s = 2$) and 6 MeV ($s = 4$) for neutrons and 3 MeV ($s = 2$) and 7 MeV ($s = 4$) for protons are ascribed to the contributions of corresponding one-pair-broken and two-pair-broken states.

\begin{figure}[ht!]
\centering
 \includegraphics[scale=.6]{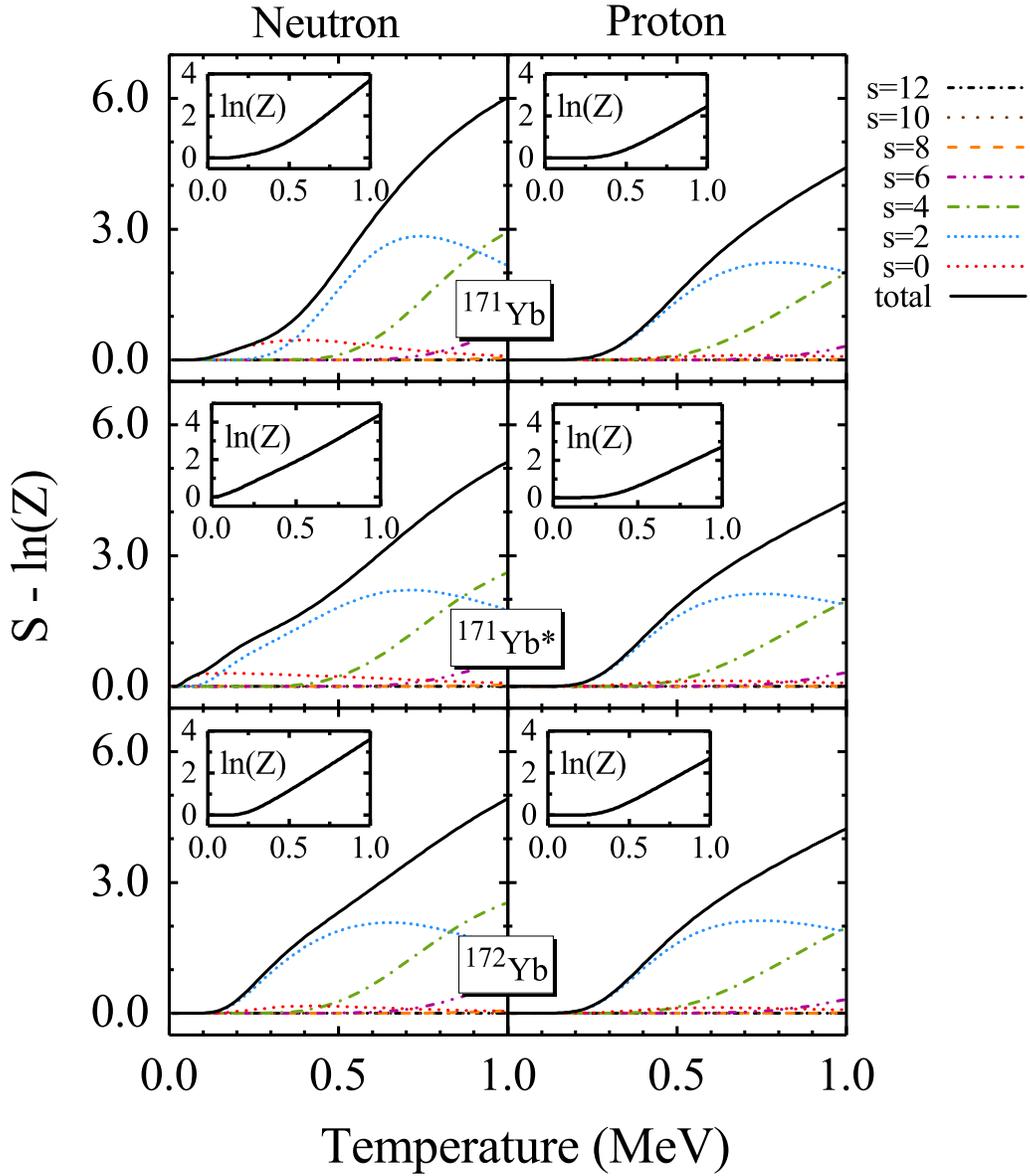}
 \caption{(Color online) Neutron (the left column) and proton (the right column) entropy subtract by $\ln Z$ for $^{171}$Yb (the top row), $^{171}$Yb* (the middle row), and  $^{172}$Yb (the bottom row) for different seniority numbers $s = 0,2,4,6,8,10,12$ and the total contribution as functions of temperature. Insert: $\ln Z$ as functions of temperature.}
 \label{fig:entropy}
\end{figure}

With the help of defined seniority components, the thermodynamic properties of hot nuclei can be studied as well. The equation of entropy is shown in Eq.~(\ref{eq:s}). According to the definition of the seniority component, the average energy can be decomposed into contributions of different seniority components easily. Then we can study the contributions of different seniority states to the entropy of hot nuclei. In Fig.~\ref{fig:entropy}, the neutron and proton entropy subtracted by $\ln Z$ for $^{171}$Yb (the top row), $^{171}$Yb* (the middle row), and $^{172}$Yb (the bottom row) with different seniority numbers $s = 0,2,4,6,8,10,12$ and the total contribution as functions of temperature are illustrated. The curves of $\ln Z$ for neutrons and protons as functions of temperature are also shown in the insert figures. It can be found that the total entropy minus $\ln Z$ for neutrons of $^{171}$Yb and $^{172}$Yb are zero at $T \le 0.1$ MeV and $T \le 0.17$ MeV respectively, while that of $^{171}$Yb* is almost never zero up to $T = 1$ MeV. The total entropy minus $\ln Z$ for protons is zero at $T \le 0.2$ MeV. For neutrons of $^{171}$Yb and $^{172}$Yb, the $s = 2$ states contribute at $T > 0.3$ MeV and $T > 0.17$ MeV respectively, and for protons of $^{171}$Yb, $^{171}$Yb*, and $^{172}$Yb, the s = 2 states contribute at $T > 0.2$ MeV. For all of the curves, the $s = 4$ states appear at $T > 0.45$ MeV. However, other higher seniority states do not contribute to the entropy up to $T = 1$ MeV. It can be understood by the fact that the $s = 0$ states do not absorb any energy. Particles just undergo transitions among single-particle levels. However, for those $s \neq 0$ states, the entropy increases with respect to temperature because they need extra energy to break particle pairs. These procedures are irreversible. Due to the limitation of model space, higher seniority contributions to the entropy are not presented up to 1 MeV.

\section{Summary}
\label{sec:sum}
In summary, the hot nuclei $^{171}$Yb and $^{172}$Yb have been investigated by the CDFT with PC-PK1 effective interaction. The pairing correlations of odd-A and even-even nuclei have been treated by the uniform framework, namely the shell-model-like approach in which the particle numbers are conserved exactly and in which the blocking effect is treated accurately. The thermodynamic quantities have been evaluated in the canonical ensemble theory. 

It is found that the one-pair-broken states and two-pair-broken states aplay crucial roles in the appearance of the S shape of the heat capacity curve in $^{172}$Yb. Moreover, due to the effect of the particle-number conservation, our calculation of the pairing gap indicates a gradual transition from the superfluid to the normal state.
The calculation results of the pairing gap in odd-A nuclei indicate the weak transition of pairing correlations from the superfluid state to the normal state in the hot nucleus compared to the even-even nucleus.
This shows that the S-shaped heat capacity curve of $^{171}$Yb does not correspond to the pairing transition.
However, the CDFT results show that the single-particle level density for neutrons near the Fermi surface of $^{171}$Yb is much smaller than that of $^{171}$Yb*.
A relatively larger energy gap has been found above and below the Fermi surface of neutrons for $^{171}$Yb.
Therefore, when the same temperature is raised, $^{171}$Yb is more difficult to generate pairing excitation.

We also find that it is very crucial to correctly handle the blocking of the single-particle levels near the Fermi surface when studying properties of odd-A nuclei. The blocking effect gives an explanation to the almost linear increasing heat capacity of $^{171}$Yb*.
Although the contributions of different seniority states will still show S-shaped curves, it is suppressed by the blocking effect in $^{171}$Yb*. However, due to the influence of the single-particle level structure, the S-shaped heat capacity curve of $^{171}$Yb appears, which should have been smoothed out by the blocking effect.

\begin{acknowledgments}
This work was supported by National Natural Science Foundation of China under Grant No. 11775099. P. W. Zhao and Z. H. Zhang are gratefully acknowledged for reading the manuscript.
\end{acknowledgments}

\bibliographystyle{apsrev4-1}
\bibliography{heat}

\end{document}